\documentstyle[epsf]{europhys}

\newif\ifboo \boofalse

\def\Review#1{\boofalse{\it #1},}
\def\Name#1{{\sc #1},}
\def\Vol#1{\ifboo Vol. {\bf #1}\else{\bf #1}\fi}
\def\Year#1{\ifboo #1\else(#1)\fi}

\def\Page#1{\ifboo {\rm p. #1}\else{\rm #1}\fi}

\begin{document}
\newpage
\euro{}{}{}{}
\Date{ }
\shorttitle{K. Kang PARITY EFFECT IN A MESOSCOPIC SUPERCONDUCTING RING}

\title{Parity Effect in a mesoscopic superconducting ring}

\author{Kicheon Kang\inst{1}\footnote{Electronic address:
        kckang@phy0.chonbuk.ac.kr}}
\institute{
    \inst{1} Department of Physics, Chonbuk National University,
             664-14 1-ga Duckjin-Dong, Duckjin-Gu
             Chonju, Chonbuk 561-756, South Korea}
\pacs{
\Pacs{74}{20Fg}{BCS theory and its development}
\Pacs{73}{23Ra}{Persistent currents}
\Pacs{73}{23$-$b}{Mesoscopic systems}
      }


\maketitle

\begin{abstract}
We study a mesoscopic superconducting ring threaded by a magnetic
flux when the single particle
level spacing is not negligible.
It is shown that, for a superconducting ring with even parity,
the behavior of persistent current is equivalent to what is 
expected in a bulk superconducting ring. On the other hand, we find
that a ring with odd parity shows anomalous behavior such as
fluxoid quantization at half-integral multiples of the flux quantum
and paramagnetic
response at low temperature. We also discuss how the parity effect
in the persistent current disappears as the temperature is raised or
as the size of the ring increases.
\end{abstract}
%
%
What happens to superconductivity when the sample is made
very small? Anderson~\cite{anderson59} already addressed this fundamental
question in 1959 and argued that as the 
size of a superconductor decreases and, accordingly,
the average level spacing $\delta$ becomes larger than the
BCS gap $\Delta$, superconductivity is no longer possible.
Recent experiments on ultrasmall ``superconducting"
nanoparticles~\cite{ralph} have led to reconsider 
this old, but fundamental question. 
In a series of experiments the authors of \cite{ralph}
 studied transport through
nanometer-scale Al grains and succeeded to get the discrete eigenspectrum
of a single superconducting grain.  The results were found to
depend on the parity, i.e. on the electron number in the grain
being even or odd. These experiments initiated several theoretical
investigations. von Delft {\em et al.}~\cite{delft96}
used a model of uniform level spacing in a parity-projected mean field
theory~\cite{janko94} and found that the breakdown of superconductivity
occurs at a value of $\delta/\Delta \sim O(1)$ which is parity-dependent.
This parity effect has been shown to increase when including the effects
of level statistics~\cite{smith96}. Effects of quantum 
fluctuations~\cite{matveev97}, canonical description of BCS 
superconductivity~\cite{mastellone98}, as well as transport theory for
a nanoparticle coupled to superconducting leads~\cite{kang98}
have also been subjects of
the study along this line. 

Another interesting example for studying the size effect on
superconductivity is a mesoscopic superconducting ring
threaded by a magnetic flux $\Phi$. 
It is well known that a conventional BCS 
superconducting ring exhibits fluxoid quantization at 
integer multiples of the flux quantum $\Phi_0=hc/2e$ and
a diamagnetic response at $\Phi=n\Phi_0$ with $n$ being an 
integer~\cite{tinkham96}.
In this Letter, we address the following question which is essentially
the same as in a simply connected grain: 
What happens in a superconducting 
ring when the size of the ring becomes very small? For this purpose,
we adopt the parity-projected mean field theory~\cite{janko94}
for an ideal mesoscopic ring.
It is shown that the order parameter 
strongly depends on the parity when the size of the ring is small
enough, as in the case of a grain. The most dramatic feature 
we show in our study
is the behavior of the supercurrent (or persistent current) 
which strongly depends on the parity. For a ring with even parity,
the behavior of the supercurrent is identical to that of
a bulk superconducting ring. It exhibits fluxoid quantization at
integer multiples of the flux quantum, and a diamagnetic 
response for small deviations
of the flux from the integer multiples of the flux quantum. 
On the other hand, the characteristics are found to be
very different for
a small, odd parity ring. In a superconductor with odd parity,
there is one unpaired quasiparticle. We find that at low
temperature the existence
of this quasiparticle drives the superconducting ring to a {\em
half-integral} fluxoid quantization and a paramagnetic
response at small values of the flux.
We further show that the
anomalous behavior in an odd-parity superconductor
disappears at temperatures 
higher than the level spacing of single electron spectra, 
where one recovers the behavior of a conventional superconducting ring.
Finally, this 
parity-dependent behavior of 
supercurrent is shown to disappear in the thermodynamic limit.

The ideal 
superconducting ring can be described by the Hamiltonian:
\begin{equation}
 H = \sum_{j\sigma}\varepsilon_j^0 c_{j\sigma}^{\dagger}c_{j\sigma} 
   -\lambda\delta \sum_{i,j} 
     c_{i\uparrow}^{\dagger}c_{\bar{i}\downarrow}^{\dagger}
     c_{\bar{j}\downarrow}c_{j\uparrow} .
\end{equation}
The single particle energy is given by
$\varepsilon_j^0=\frac{\hbar^2}{2mR^2}(j-f/2)^2$, where 
$R$ is the radius of the ring,
$f=\Phi/\Phi_0$ is the external
flux divided by the flux quantum $\Phi_0=hc/2e$, and
$j$ is an integer which corresponds to an angular momentum quantum number.
This is obtained by solving the Schr\"odinger equation in a
1D noninteracting ring.    
Note that $\bar{j}=-j$.
$\lambda$ is the dimensionless BCS coupling constant.
$\delta=\hbar^2N/8mR^2$ is
the level spacing at the Fermi energy,
where $N$ is the number of electrons in the ring.
We don't take into account the Zeeman splitting, namely $h$,
because it is negligible unless the radius of the ring is very small.
For $\Phi\sim\Phi_0$ with a uniform magnetic field,
the ratio of the level spacing $\delta$ to the Zeeman
splitting is proportional to $R$ which is
estimated as $\delta/h\sim 10^{20} r_s R / m^2$, where $r_s$ is the 
average distance between electrons. 
For example, in a typical superconductor such as Al with $R\sim 1\mu m$,
$\delta/h\sim 10^4$.

A simple way of describing a mesoscopic superconductor 
with fixed number parity $P$ (denoted by $e$ for even, 
and $o$ for odd parity) is to adopt 
the parity-projected grand canonical partition function~\cite{janko94} 
\begin{equation}
 Z_P(\mu) = \mbox{\rm Tr} \frac{1}{2}[ 1\pm (-1)^N ] e^{-\beta(H-\mu N)} .  
\end{equation}
We evaluate $Z_P$ using the BCS-type 
mean field approximations, which consists in 
neglecting quadratic terms of the fluctuations :
\begin{eqnarray}
 & & c_{i\uparrow}^{\dagger}c_{\bar{i}\downarrow}^{\dagger}
     c_{\bar{j}\downarrow}c_{j\uparrow} \nonumber \\ 
    &\simeq& 
 \langle c_{i\uparrow}^{\dagger}c_{\bar{i}\downarrow}^{\dagger}\rangle
 c_{\bar{j}\downarrow}c_{j\uparrow} 
     +
 c_{i\uparrow}^{\dagger}c_{\bar{i}\downarrow}^{\dagger}
 \langle c_{\bar{j}\downarrow}c_{j\uparrow} \rangle
     -
 \langle c_{i\uparrow}^{\dagger}c_{\bar{i}\downarrow}^{\dagger} \rangle
 \langle c_{\bar{j}\downarrow}c_{j\uparrow} \rangle \label{eq:mean} \\
    &+& \delta_{ij} \left(
 c_{i\uparrow}^\dagger c_{i\uparrow}
 \langle c_{\bar{i}\downarrow}^\dagger c_{\bar{i}\downarrow} \rangle
     +
 \langle c_{i\uparrow}^\dagger c_{i\uparrow} \rangle
 c_{\bar{i}\downarrow}^\dagger c_{\bar{i}\downarrow}
     -
 \langle c_{i\uparrow}^\dagger c_{i\uparrow} \rangle
 \langle c_{\bar{i}\downarrow}^\dagger c_{\bar{i}\downarrow} \rangle
    \right) . \nonumber
\end{eqnarray} 
The ensemble average $\langle\cdots\rangle$ should be
evaluated in a given parity $P=e$ or $P=o$.
The first three terms on the r.h.s of Eq.(\ref{eq:mean}) correspond to
the mean-field approximation for the superconducting 
pairing. The last three terms are usually
not considered since they give no
contribution in the thermodynamic limit.  However, 
those terms cannot be ignored in mesoscopic systems,
though they were
neglected in the previous mean-field description for ultrasmall
superconducting grains~\cite{delft96}. 
Note that the validity of our mean-field treatment is limited to
the $\delta < \Delta$ limit where the usual BCS approximation can
be applied.
As a result we get the following expression for the 
mean-field Hamiltonian 
\begin{equation}
 H = C_P + \sum_{j\sigma} 
          \tilde{E}_j\gamma_{j\sigma}^\dagger \gamma_{j\sigma} 
   + \mu N \; ,
 \label{eq:mfh}
\end{equation}
where $\gamma_{j\sigma}$ ($\gamma_{j\sigma}^\dagger$)
destroys (creates) a quasiparticle; $\gamma_{j\sigma}=u_j c_{j\sigma}-\sigma 
v_j c_{\bar{j}\sigma}^\dagger$,
and the constant $C_P$ and the quasiparticle energy $\tilde{E}_j$
are given by
\begin{eqnarray}
 C_P &=& \sum_j \left[ \frac{1}{2}
       (\varepsilon_j+\varepsilon_{\bar{j}}) - E_j
              \right]
     + \Delta_P^2/\lambda\delta  \\
     &+& \lambda\delta \sum_j \left[ u_j^2 f_{\bar{j}\bar{\sigma}}+
       v_j^2(1-f_{j\sigma}) \right]
       \left[ u_j^2 f_{j\sigma}+
       v_j^2(1-f_{\bar{j}\bar{\sigma}}) \right] , \nonumber
\end{eqnarray}
and 
\begin{equation}
 \tilde{E}_j = \frac{1}{2}
    (\varepsilon_j-\varepsilon_{\bar{j}}) + E_j ,
 \label{eq:quasi}
\end{equation}
respectively.
Here $\Delta_P=\lambda\delta \sum_j\langle c_{\bar{j}\downarrow}c_{j\uparrow}
\rangle$ is the parity-dependent order parameter which 
has to be calculated
self-consistently.
$\varepsilon_j$ and $E_j$ are defined as $\varepsilon_j=\varepsilon_j^0-\mu
-\lambda\delta [u_j^2f_{\bar{j}\bar{\sigma}}+v_j^2(1-f_{j\sigma})]$
and $E_j=\sqrt{\Delta_P^2+(\varepsilon_j+\varepsilon_{\bar{j}})^2/4}$,
respectively.
For simplicity we neglect the last term of $\varepsilon_j$. 
It does not change the qualitative feature of our results since its
role is only to increases somewhat the effective
level spacing near the Fermi level for large $\delta$.
$f_{j\sigma} = \langle\gamma_{j\sigma}^\dagger\gamma_{j\sigma}\rangle$
is the average occupation number of a state $j$ with
spin $\sigma$. It is parity-dependent and given by
\begin{equation}
 f_{j\sigma} = \frac{ f_+(\tilde{E}_j)Z_+ \mp f_-(\tilde{E}_j)Z_- }{
                      Z_+\pm Z_- } \;\; ,
\end{equation}
for $P=e$ (upper sign) and $P=o$ (lower sign), where
$f_{\pm}(E) = 1/(e^{\beta E}\pm 1)$ and 
$Z_\pm = \prod_{j\sigma} ( 1\pm e^{-\beta\tilde{E}_j} )$.
$u_j,v_j$ are BCS coherence factors: 
\begin{equation}
 u_j^2 = 1-v_j^2 = \frac{1}{2} 
 \left( 1+\frac{\varepsilon_j+\varepsilon_{\bar{j}}}{ 2E_j } \right) .
\end{equation}
The parity-dependent chemical potential $\mu_P$ is determined by
the relation
\begin{equation}
 \langle N\rangle = \frac{1}{\beta} \frac{\partial}{\partial\mu}
  \log{Z_P(\mu)}|_{\mu=\mu_P}   \;\; ,
\end{equation}
that holds 
provided that $\mu_P$ lies halfway between the last filled and
first empty level if $P=e$, and on the singly occupied level if 
$P=o$~\cite{delft96}.

The mean-field self-consistency condition 
$\Delta_P=\lambda\delta \sum_j\langle c_{\bar{j}\downarrow}c_{j\uparrow}
\rangle$ leads to the relation
\begin{equation}
 1 = \lambda\delta \sum_{j=-j_c}^{j_c} 
  \frac{1}{2E_j} (1-f_{\bar{j}\downarrow}-f_{j\uparrow})
 \; ,
 \label{eq:gap}
\end{equation}
where $j_c$ is the cutoff value of $j$ in the summation.
At $T=0$ with $\Phi=0$  the occupation of quasiparticles reduces to
$f_{j\sigma} = \frac{1}{4}$ if the level lies on the chemical potential,
namely $j=\pm j_F$, for $P=o$, and zero otherwise.
The factor $1/4$ is due to the orbital degeneracy of the ring,
$\varepsilon_j^0=\varepsilon_{\bar{j}}^0$,
as well as the spin degeneracy. For nonzero external flux, the orbital
degeneracy is lifted and 
$f_{j_F\sigma}=1/2$ (but $f_{\bar{j}_F\sigma}=0$) for odd parity.
This nonzero value of quasiparticle occupation gives rise to the so
called ``blocking effect". That is, the odd parity superconductor has
one unpaired electron, which prevents pair scattering of other pairs
into/out of the singly occupied state
and reduces the order parameter as compared to that of an even
parity superconductor\cite{braun99}.

In the framework of the parity-projected grand canonical
description, three different cases appear in solving the ``gap"
equation:
(i) Even parity with fully occupied highest level ($N=4j_F+2$),
(ii) even parity with partially occupied highest level
($ N = 4j_F$), and
(iii) odd parity ($N = 4j_F \pm 1$).
It is important to note that the self-consistent equation (\ref{eq:gap})
does not depend 
on the flux if $\langle N\rangle$ is kept unchanged under variation of
the flux. The condition of constant $\langle N\rangle$ leads to
the flux-dependent chemical potential 
\begin{equation}
 \mu_P(f) = \mu_P(0) + \frac{\hbar^2}{8mR^2} f^2.
 \label{eq:mu}
\end{equation}  
$E_j$ is independent of $f$ with this condition, and accordingly
Eq.(\ref{eq:gap}) is also flux-independent at low temperature,
$k_BT < \delta, \Delta_P$.

\begin{figure}
\epsfxsize=4.5in
\epsffile{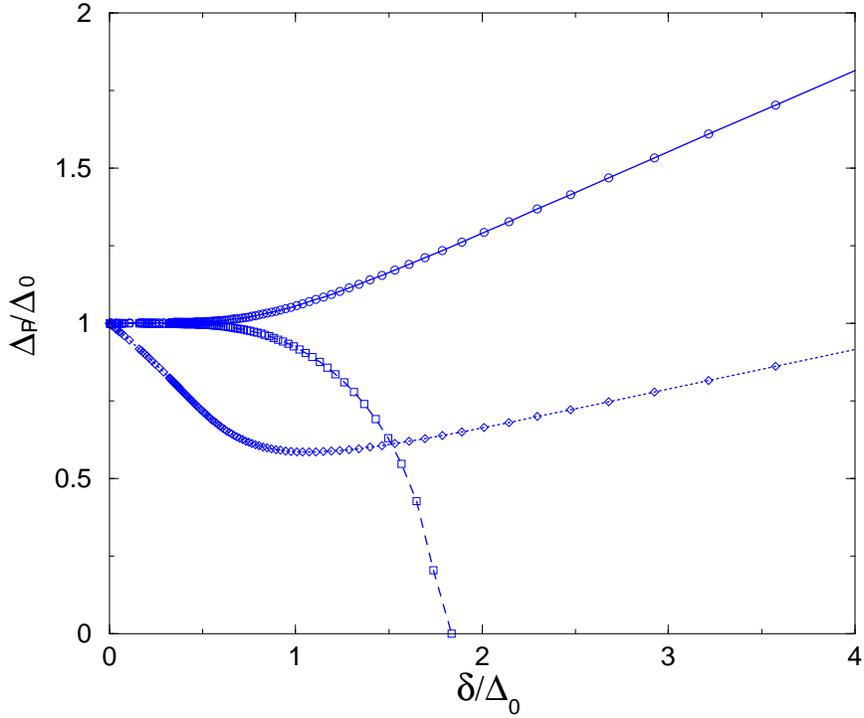}
 \caption{ The pairing parameters as a function of the single particle
           level spacing 
           for $\lambda=0.2$ with $j_c=2j_F$.
           Lines with symbols show the results for three different cases,
           $P=e$ with $N=4j_F$ (solid line with
           circles), $P=o$ with $N=4j_F\pm1$ (dotted line with
           diamonds), and $P=e$ with $N=4j_F+2$ (dashed line
           with squares). $\Delta_0$ denotes the bulk gap. 
	   }
 \label{fig:gap}
\end{figure}

Fig.\ref{fig:gap} shows the pairing parameter 
as a function of $\delta$
keeping the electron density
constant. In solving the equation we chose $\lambda=0.2$, close to
that of Al~\cite{carbotte90}, and $j_c=2j_F$. 
The pairing parameter which is obtained by solving Eq.(\ref{eq:gap})
depends strongly on the parity as in the grain superconductor.
In the odd-parity superconductor the order parameter
is suppressed compared to the one for even parity.
However,
the behavior of the pairing parameter
as a function of the level spacing is somewhat
different from the grain superconductor with equal level
spacing~\cite{delft96} or with randomly distributed levels~\cite{smith96}.
It is due to the existence of orbital degeneracy in the ideal
ring geometry.
If the highest occupied level is fully filled (case (i)), 
the chemical potential
lies halfway between the last filled and the first empty level.
In this case there exists a critical level spacing $\delta_c/\Delta_0
\sim O(1)$ where
$\Delta_e$ goes to zero as in the mean-field solution for the 
grain superconductivity.
If the highest occupied level is partially filled
(both for even and odd parity) the chemical potential lies on that
level. Pair scattering of electrons between these orbitally degenerate
levels makes it impossible to have a solution with $\Delta_P=0$. 
On the other hand, the presence of any weak disorder will
lift the orbital degeneracy and give a solution with $\Delta_P=0$.
Note that, in a very small grain with $\delta\gg\Delta_0$,
quantum fluctuations are important
and the mean-field treatment becomes invalid~\cite{matveev97}.

Next we discuss the parity-dependent behavior of the persistent current 
in the ``superconducting" state.
Because we deal with an isolated ring, the current should be calculated
in the canonical description, while we use the
parity-projected grand canonical partition function. 
However, the number fluctuation
in the grand canonical treatment
is very small at low temperature and the free energy with a fixed number
$N$, namely $F_N$, can be written as
\begin{equation}
 F_N \simeq -\frac{1}{\beta} \log{Z_P} + \mu_P N .
\end{equation}
Note that this relation is exact at $T=0$.
The persistent current is given by  
\begin{equation}
 I = -c\frac{\partial F_N}{\partial\Phi} ,
\end{equation}
which gives the following expression 
by using the mean field Hamiltonian (\ref{eq:mfh}) 
\begin{equation}
 I = I_{dia} + I_{para} ,
\end{equation}
where $I_{dia}$ and $I_{para}$ are the diamagnetic and the
paramagnetic contribution
to the current, respectively,
\begin{eqnarray}
 I_{dia} &=& -c\frac{\partial\mu_P}{\partial\Phi} N \; , \label{eq:dia}\\
 I_{para} &=& -c\sum_{j\sigma} 
   \frac{\partial\tilde{E}_j}{\partial\Phi} f_{j\sigma}
  \;\;. \label{eq:para}
\end{eqnarray}
Note that at low temperature ($k_BT < \delta$, $\Delta_P$) 
the constant term $C_P$ 
in the Hamiltonian (\ref{eq:mfh}) does not contribute
to the current since it is independent of the flux.
 
\begin{figure}
\epsfxsize=4.5in
\epsffile{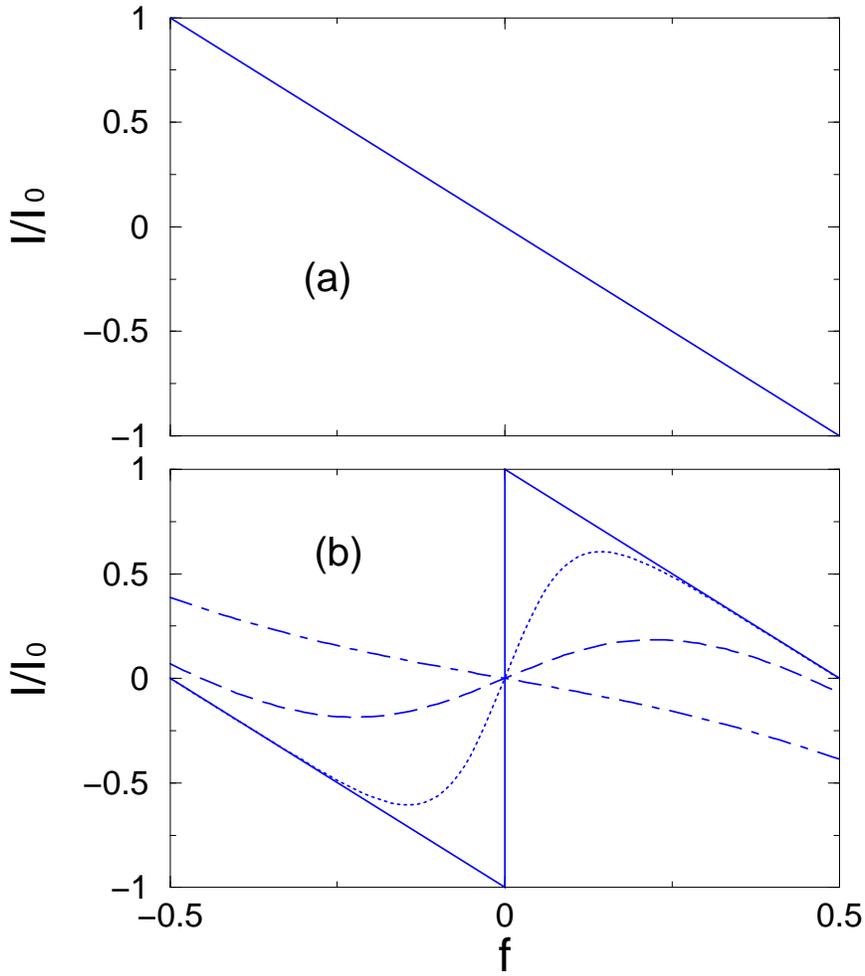}
 \caption{ The persistent current
           as a function of the dimensionless Aharonov-Bohm flux 
           $f=\Phi/\Phi_0$ with $\Phi_0=hc/2e$ (a) for even parity, and
           (b) for odd parity at $t=0$ (solid line), $t=0.1$
           (dotted line), $t=0.3$ (dashed line), and $t=0.7$
           (dot-dashed line).
           }
 \label{fig:current}
\end{figure}

For even parity, the paramagnetic contribution to the persistent current
is absent at $k_BT\ll\Delta_e$ since $f_{j\sigma}=0$ 
for $P=e$. Thus, the persistent
current for $P=e$, $I^e$, is equal to $I_{dia}$ 
and obtained from Eq.(\ref{eq:mu}) and (\ref{eq:dia}):
\begin{equation}
 I^e = -2I_0 f \;\; ,
\end{equation}
where $I_0=ev_F/(2\pi R)$ with $v_F=\frac{\hbar}{m}\frac{j_F}{R}$ 
being the Fermi velocity and $f=\Phi/\Phi_0$.
(See Fig.\ref{fig:current}(a).)
The behavior of $I^e$ is 
essentially equivalent to what is expected in the bulk
superconducting ring.  

On the other hand, the paramagnetic contribution is important
for $P=o$ because of an unpaired quasiparticle. $I_{dia}$ is identical
for $P=o$ and $e$. At low temperature
($k_BT < \delta$) $I_{para}$ can be written as follows 
\begin{equation}
 I_{para} \simeq 2I_0 \left( f_{j_F\sigma}-f_{\bar{j}_F\sigma} \right) ,
\end{equation} 
where 
\begin{equation}
 f_{\pm j_F\sigma} \simeq \frac{1}{2} \frac{ e^{\pm f/t} }{
   \left( e^{f/t} + e^{-f/t} \right) } \;\; . 
\end{equation}
$t$ is dimensionless temperature 
$t = k_BT/\delta$.
At zero temperature $I_{para}$ reduces to $\mbox{\rm sgn}
[f] I_0$ and the total
current for $P=o$, $I^o$, is given by
\begin{equation}
 I^o = I_0(\mbox{\rm sgn} [f] -2f)  \; .
\end{equation}
As shown in the solid line of Fig.\ref{fig:current}(b),
the current vanishes at half-flux quantum, $f=\pm 1/2$,
while it has maximum value of $|I^o|$
at $f=0$. This implies that the possible 
values of fluxoid in a ring with odd parity are half-integers,
$\pm\Phi_0/2$, $\pm3\Phi_0/2$, etc, in contrast to the conventional
fluxoid quantization at integer multiples of $\Phi_0$. 
Note that, the fluxoid $\Phi'$ 
is defined in a conventional way~\cite{tinkham96}
\begin{equation}
 \Phi' = \Phi + \frac{(2m)c}{2e} \oint v_s\cdot dl \;\; ,
\end{equation}
where $v_s$ is the supercurrent velocity.

The parity-dependent behavior of the persistent current
found above should be distinguished from that of a normal ring
of noninteracting electrons with spin $1/2$ considered by Loss
and Goldbart.~\cite{loss91} In a normal ring with spin $1/2$, the
persistent current depends on the number of particles with modulo
4. Periodicity and oscillation amplitude depend on the number, and
the current response is diamagnetic only for $N=4j_F+2$ (the case
of fully occupied top level). On the other hand, 
a mesoscopic superconducting ring shows only the parity dependence
that originates from the electron paring. The periodicity and amplitude
of oscillation remain unchanged as $\Phi_0 (=hc/2e)$ and $I_0$, 
respectively.  

The effect of temperature on the persistent current is shown in
Fig.\ref{fig:current} (b). $I_{para}$ is 
reduced as the temperature is raised. 
It is clear that the effect of the unpaired quasiparticle disappears
with increasing $T$; the persistent current will show a behavior 
like in Fig.\ref{fig:current}(a), 
at temperature higher than $t\sim 1$.
Actually there is a crossover
temperature $t^*=0.5$ where the current response at small flux
changes from paramagnetic to diamagnetic.
 (That is $\left.\frac{dI^o}{d\Phi}\right|_{\Phi=0}
 (t=t^*)=0$.)

Let us, finally, discuss the condition under which the parity effect
can be observed in the persistent current.
As noted above, the condition for the existence of the paramagnetic
contribution to the persistent current is $T < T^*$
with $k_BT^*/\delta=0.5$. 
Thus, the crossover temperature is given by 
\begin{equation}
 k_BT^* = 2.5 \mbox{\rm (eV)} \frac{r_s}{R} \;\;,
\end{equation}
where $r_s$ is the average distance between electrons.
For Al $r_s=1.10$\AA$=2.07a_0$, with $a_0$ being the Bohr radius. 
For a ring with $R\sim 5\times 10^4r_s\sim 5\mu m$, 
$T^*\sim 1\mbox{\rm K}$ 
is comparable to the transition temperature of bulk Al. 
If the ring is not strictly 1D, the level spacing is reduced
and, accordingly, the crossover temperature
$T^*$ is lowered as compared to the ideal 1D ring with the 
same radius. 
Furthermore, $T^*\rightarrow 0$ if $R\rightarrow\infty$. 
In other words, the parity effect disappears in the thermodynamic
limit, and our treatment coincides with the 
conventional BCS description.

In conclusion, we have investigated 
the parity-dependent properties of 
a mesoscopic superconducting ring threaded by a magnetic
flux. 
The properties in the persistent current
of an even-parity ring are similar to those of a bulk
superconductor. On the other hand, a ring with odd parity shows
unconventional behavior: the fluxoid quantization at {\em half}-integer
multiples of the flux quantum. Moreover, there exists a paramagnetic
response at temperatures below a characteristic temperature of the order
of the level spacing. 
We have also shown that this parity effect of the persistent current 
disappears as the temperature is raised or
as the size of the ring increases.
 
The author wishes to acknowledge valuable discussions with A. Bill.
This work has been supported by the Visitors Program of the MPI-PKS,
and also by the BK21 project of the Korean Ministry of Education.


%
%
\end{document}